\documentclass[10pt,twocolumn]{article}

\usepackage{amssymb}
\usepackage{amsmath}
\usepackage{graphicx}

\begin{document}

\title{Acceleration in de Sitter spacetimes}

\author{Ion I. Cot\u aescu \\
 {\small West University of Timisoara} \\{\small V. Parvan Ave. 4, RO-300223 Timisoara, Romania}}
{\onecolumn
\maketitle

\abstract{We propose a definition of  uniform accelerated frames in de Sitter spacetimes applying  the Nachtmann method of introducing coordinates using suitable point-dependent isometries.  In order to recover the well-known Rindler approach  in the flat limit, we require the transformation between the static frame and the accelerated one to depend continuously on acceleration, obtaining  thus the natural generalization of the Rindler transformation to the de Sitter spacetimes of any dimensions.

Pacs: 04.20.Cv, 04.62.+v}

 }

\newpage
\twocolumn
\section{Introduction}

The problem of  accelerated motions in de Sitter expanding universes is of a special  interest for understanding the effects of acceleration in the presence of the natural expansion. A general method in investigating these effects relies on conformal transformations among different metrics. Thus in Ref. \cite{PG} the metrics of the static and accelerated de Sitter frames are related by a  conformal scaling depending on acceleration. Another possibility  is to consider some special frames of the de Sitter and Minkowki spacetimes whose metrics are conformal with the metric of the Einstein universe \cite{BK1,BK1a,BK2}. In this manner one can exploit the conformal invariance of the field equations obtaining important results concerning the scalar or electromagnetic radiation produced by the accelerated scalar or electric charges in de Sitter spacetimes.

In the present Letter we would like to propose a different approach trying to construct a natural generalization of the Rindler transformations to the de Sitter spacetime. The Rindler theory of  uniform accelerated frames in Minkowski spacetime \cite{MB,R} is the geometric framework in which the Unruh effect was introduced and then intensively studied  \cite{unruh,crispino,BD,unruh_dw,fedotov,takagi,fulling,lonsol}.  
In other respects, the Rindler metric can be formally related to a chart of  the de Sitter manifold \cite{BD, SS} whose metric is a conformal Rindler one. We may ask if could we interpret this chart as being  of an accelerated observer in the de Sitter spacetime or as a mere mathematical coincidence.  Our aim is to solve this problem showing that by generalising the Rindler transformation  to the de Sitter geometry we obtain a new specific metric of the (uniform) accelerated frames on this manifold.

In the two-dimensional Minkowski spacetime the Rindler transformation is defined up to a space translation since the translations are Abelian isometries of this space. However, if we intend to generalize the Rindler approach to de Sitter spacetimes, we must take into account that there the  space translations are not commuting with the time ones \cite{C1}. For this reason, we start with fixed initial conditions restricting ourselves to the {\em continuous} transformations with respect to the acceleration $a$, in $a=0$. Therefore, the coordinates of  the static frame $\{T,X\}$ and those of the accelerated one $\{t,x\}$ (with $x\in {\Bbb R}$ and $t\in{\Bbb R}^+$)  have to transform as     
\begin{eqnarray}
T&=&\frac{e^{a x}}{a} \sinh(a t)\,,\label{T}\\
X&=&\frac{e^{a x}}{a} \cosh(a t)-\frac{1}{a}\,,\label{X}
\end{eqnarray} 
since then the limits $T\to t$ and $X\to x$ for $a\to 0$ guarantee the continuity  in $a=0$.  The accelerated frame, along the positive $x$ axis ($a>0$), covers the right-hand Rindler wedge where $X\in [-a^{-1},\infty)$, while for $a<0$  the left-hand wedge is its symmetric with respect to $X=0$. We note that, in this chart, the Rindler metric  is independent of  translations taking  the well-known form \cite{BD},  
\begin{equation}\label{dsR}
ds^2=dT^2-dX^2=e^{2ax}(dt^2-dx^2)\,,
\end{equation}
which  behaves continuously in $a=0$.   In what follows,  we  restrict ourselves to study only the case of positive accelerations in the right-hand wedge. 

We must specify that the translation introduced in Eq. (\ref{X}), for assuring the continuity  in $a=0$, determines the identity $i\partial_t=i\partial_T+a i(T\partial_X+X\partial_T)$ that in a self-explanatory notation can be written in terms of quantum observables as $H_{acc}=H+aK$, where $H$ is the energy operator and $K$ the Lorentz generator of the static frame.  However,  in the case of the symmetric Rindler transformation \cite{BD}  which is singular in $a=0$ (without translating $X$ with $-a^{-1}$) this identity reduces to $H_{acc}=a K$. Recently we  found a  metric of a mobile frame in de Sitter spacetime whose energy operator  is just $aK$, but this model seems to not have a satisfactory physical meaning \cite{Cprep}. This is another argument for considering here the continuous  Rindler transformations (\ref{T}) and (\ref{X}).

Our principal purpose  is to generalize this conjuncture to the de Sitter manifolds  requiring the transformation between the static and accelerated frames (I) to be continuous in $a=0$ and (II)  to recover the above Rindler transformation in the flat limit. We first present our proposal in the simpler case of two-dimensional de Sitter manifolds, following thereafter to  generalize it to the de Sitter manifolds with arbitrary dimensions. 

The method we use here was proposed by Nachtmann for constructing  covariant representations of the fields defined on  the de Sitter spacetime, seen as a homogeneous space of its own  isometry group \cite{Nach}.  The idea is to apply the Wigner method  of  orbital analysis,  but in configurations instead of the momentum representation. In this manner various systems of coordinates can be introduced by choosing suitable point-dependent isometry transformations which are called here {\em boosts} \cite{C2}. 

The paper is organized as follows. In the second section we present the method of boosting coordinates on de Sitter manifolds, pointing out how  the canonical 1-forms and the components of the Killing vector fields can be derived in any de Sitter chart. We first give the boost of the de Sitter-Painlev\'e chart \cite{G,P,Pascu} which is considered here the frame of the static (or fixed) observer.  In the next section we postulate the form of  the boosts of (uniform) accelerated frames in two dimensions and  deduce the Rindler-type transformation between these frames, showing that this complies with our requirements (I) and (II).  Moreover, investigating other properties, we find that the event horizon of the accelerated frame is pushed forward in the acceleration direction while the time-like Killing vector fields accomplish the same transformation rule as the one we met in the flat case mentioned above.  All these results are generalised to an arbitrary number of dimensions in the fourth section. Finally, we briefly present our concluding remarks. 

\section{Boosting de Sitter frames}

Let $(M_d,g)$ be the $d$-dimensional  de Sitter spacetime defined as the hyperboloid of radius $1/\omega$ \footnote{We denote by $\omega$ the Hubble-de Sitter constant since  $H$ is reserved for the energy operator.} in the $(d+1)$-dimensional flat spacetime $(\hat M_{d+1},\hat \eta)$ of coordinates $z^A$  (labelled by the indices $A,\,B,...= 0,1,2,...d$) and metric 
\begin{equation}
\hat\eta={\rm diag}(1,\,\underbrace{-1,-1,...-1,-1}_d)\,,
\end{equation}
which rises or lowers these indices. Any local chart $\{x\}=\{t,\vec{x}\}$  of coordinates $x^{\mu}$ ($\mu, \nu,...=0,1,2,...d-1$) can be introduced on $(M_d,g)$ giving a set of functions $z^A(x)$ which solve the hyperboloid equation,
\begin{equation}\label{hip}
\hat\eta_{AB}z^A(x) z^B(x)=-\frac{1}{\omega^2}\,.
\end{equation}
Notice that here we use the vector notation for the $(d-1)$-dimensional vectors $\vec{v}=(v^1,v^2...v^{d-1})$ of components $v^i$ ($i,j,k,...=1,2,...d-1$).

The  group $\hat G=SO(1,d)$  is simultaneously the orthogonal group of the metric $\hat\eta$ and  the isometry group of $(M_d,g)$  since its transformations  $z^A\to {\frak g}^{A\,\cdot}_{\cdot\,B}\,z ^B$ (or simply $z\to {\frak g} z$)   leave the equation (\ref{hip}) invariant for any ${\frak g}\in \hat G$.  In what follows we adopt the parametrization \cite{C2},
\begin{equation}\label{param}
{\frak g}(\xi)=\exp\left(-\frac{i}{2}\,\xi^{AB}\sigma_{AB}\right)\in SO(1,d) 
\end{equation}
with skew-symmetric parameters, $\xi^{AB}=-\xi^{BA}$,  and the covariant matrix-generators $\sigma_{AB}$ of the fundamental representation of the $so(1,d)$ algebra carried by $(\hat M_{d+1},\hat\eta)$. These have the matrix elements, 
\begin{equation}
(\sigma_{AB})^{C\,\cdot}_{\cdot\,D}=i\left(\delta^C_A\, \eta_{BD}
-\delta^C_B\, \eta_{AD}\right)\,.
\end{equation}
The principal $so(1,d)$ generators with physical meaning \cite{C1} are the energy $\hat h=\omega\sigma_{0d}$, angular momentum  $\hat j_{kl}=\sigma_{kl}$ ($i,j,k,...=1,2...d-1$), Lorentz boosts $\hat k_i=\sigma_{0i}$, and the Runge-Lenz-type vector $\hat r_i=\sigma_{id}$. In addition, it is convenient to introduce the momentum $\hat p_i=-\omega(\hat r_i+\hat k_i)$ and its dual $\hat q_i=\omega(\hat r_i-\hat k_i)$ which are nilpotent matrices  generating two Abelian $(d-1)$-dimensional subalgebras, $tr_P$ and respectively $tr_Q$. The matrices $\hat p^i\in tr_P$ generate the space translations  on  $(M_d,g)$.  All these generators may form different bases of the algebra $so(1,d)$ as, for example, the basis $\{\hat h,\hat p_i,\hat q_i,\hat j_{il}\}$ or  $\{\hat h,\hat p_i,\hat j_{il},\hat k_i\}$ \cite{C1}. We note that the $(d-1)$-dimensional restriction, $\{j_{il},k_i\}$, of the $so(1,d-1)$ subalgebra generate the vector representation of the group $SO(1,d-1)$.

Using  the parametrization (\ref{param}) we can write down  the $SO(1,d)$ isometries  $x\to x'=\phi_{{\frak g}}(x)$ according to the rule  
\begin{equation}\label{zgz}
z[\phi_{\frak g}(x)]={\frak g}\,z(x). 
\end{equation}
For example, the transformations ${\frak g}\in SO(d)\subset SO(1,d)$  generated by $\hat j_{kl}$, are generalized $SO(d-1)$ rotations in the space of the space-like vectors $\vec{z}$. In particular, the transformations we need here are: the time translations generated by $\hat h$, as
\begin{equation}\label{gh}
{\frak g}_h(\alpha) \equiv e^{i\frac{\alpha}{\omega} \hat h}=
\left(\begin{array}{cccc}
\cosh\alpha&0 &\cdots&\sinh\alpha\\
0&1&\cdots&0\\
\vdots &\vdots&{\bf 1}&\vdots\\
\sinh\alpha&0&\cdots&\cosh\alpha
\end{array}\right)\,,
\end{equation}
the space translations, 
\begin{eqnarray}
&&{\frak g}_p(\vec{x})\equiv  e^{-i{x}^i\hat p_i}=\nonumber\\
&&
\left(\begin{array}{cccc}
1+\frac{1}{2}\omega^2\vec{x}^2&\omega x^1 &\cdots&\frac{1}{2}\omega^2\vec{x}^2\\
\omega x^1&1&\cdots&\omega x^1\\
\vdots &\vdots&{\bf 1}&\vdots\\
-\frac{1}{2}\omega^2\vec{x}^2&-\omega x^1&\cdots&1-\frac{1}{2}\omega^2\vec{x}^2
\end{array}\right)\,,  \label{gp}
\end{eqnarray}
and the particular rotations in the plane $(z^{d-1},z^d)$,  
\begin{equation}\label{gr}
{\frak g}_r(\theta)\equiv e^{i \theta \hat r_{d-1}}=\left(\begin{array}{cccc}
1&\cdots &0&0\\
\vdots &{\bf 1} &\vdots&\vdots\\
0&\cdots&\cos\theta&\sin\theta\\
0&\cdots&-\sin\theta&\cos\theta
\end{array}\right),
\end{equation}
generated by the matrix $\hat r_{d-1}$.
 
Such transformations can be used for  introducing various types of coordinates \cite{Nach,C2} since the de Sitter manifold is isomorphic to the space of left cosets $\hat G/G$. Indeed, if one fixes the point $z_o=(0,0,...0,0,\omega^{-1})^T\in (\hat M_{d+1},\hat\eta)$,  then the whole de Sitter manifold is the orbit $(M_d,g)=\{{\frak g} z_o | {\frak g}\in \hat G/G\} \subset (\hat M_{d+1},\hat\eta)$ where the subgroup $G=SO(1,d-1)$ is  the stable group of $z_o$ (i. e. ${\frak g}\,z_o=z_o\,,\, \forall {\frak g}\in G$). Then any point $z(x)\in(M_d,g)$ can be reached  applying a point-dependent boost ${\frak b}(x):\,z_o\to z(x)={\frak b}(x)z_o$ which defines the functions $z^A(x)$ of the local coordinates $x$. In fact, these boosts are sections in the principal fiber bundle on  $(M_d,g)\sim \hat G/G$ whose fiber is just the isometry group $\hat G$.    

This formalism allows one to derive the {\em canonical} $(d+1)$-dimensional 1-forms $\hat\omega(x)={\frak b}^{-1}(x)d\,{\frak b}(x)\,z_o$ 
whose components 
\begin{equation}
\hat\omega^{\hat\alpha}(x)=\hat e^{\hat\alpha}_{\mu}(x)dx^{\mu}\,,\quad 
\hat\omega^{d}(x)=0\,,
\end{equation}  
give the canonical gauge  fields (i. e. tetrads or vierbeins in the case of $d=4$)  $\hat e^{\hat\mu}$ of the local co-frames associated to the fields $e_{\hat\mu}$  of the orthogonal local frames \cite{Nach}. These fields are  labelled by the local indices $\hat\alpha,\hat\mu,...$ having the same range as the natural ones. Then, the de Sitter line element of the chart $\{x\}$ can be written as  
\begin{equation}
ds^2=\hat\eta_{AB}\hat\omega^A\,\hat\omega^B=
\eta_{\hat\alpha\hat\beta}\hat e^{\hat\alpha}_{\mu}\hat e^{\hat\beta}_{\nu}dx^{\mu}dx^{\nu}=g_{\mu\nu}dx^{\mu}dx^{\nu}\,,
\end{equation} 
where $\eta={\rm diag}(1,-1,...-1)$ denotes the $d$-dimensional Minkowski metric. 

In general, the boosts are defined up to an arbitrary gauge, ${\frak b}(x)\to {\frak b}(x)\lambda^{-1}(x)$, $\lambda(x)\in G$, that does not affect the functions $z^A(x)$ but changes the gauge fields  transforming the  1-forms  as $\hat\omega(x) \to \lambda(x)\,\hat\omega(x)$    \cite{Nach,C2}. For this reason, $G$ is called the gauge group of $(M_d,g)$  associated to the metric $\eta$ of the pseudo-Euclidean model of this manifold. The gauge group $G$ is important since its representations induce the covariant representations of the isometry group $\hat G$ \cite{Nach,C1,C2}.

In other respects, it is known that starting with the functions $z^A(x)$ one can derive the components of the Killing vector fields associated to the parameters $\xi^{AB}$. This means that we may express these components in terms of matrix elements of boost transformations. Indeed, according to our previous results  \cite{C2,C1}, we find that these are $k_{(AB)}^{\mu}= 
g^{\mu\nu}\hat \eta_{AC}\hat\eta_{BD}k^{(CD)}_{\nu}$ where the components
\begin{equation}
k^{(AB)}_{\mu}(x)=\frac{1}{\omega^2}\left[{\frak b}^{A\,\cdot}_{\,\cdot\, d}(x)\stackrel{\leftrightarrow}{\partial}_{\mu} {\frak b}^{B\,\cdot}_{\,\cdot\, d}(x)\right]
\end{equation}
are written with the notation  $f\stackrel{\leftrightarrow}{\partial}_{\mu}g=f\partial_{\mu}g-g\partial_{\mu}f $.  

Hence we see that by choosing suitable boosts we may introduce various types of coordinates determining simultaneously  the gauge fields as well as the Killing vector fields. For example, the FLRW chart $\{t,\vec{x}\}$ of the expanding portion of $(M_d,g)$ is boosted by  ${\frak b}_{FLRW}(x)= {\frak g}_p(\vec{x}){\frak g}_h(\omega t)$ which gives the well-known 1-forms and the FLRW metric \cite{Nach}. If we change the positions of ${\frak g}_p$ and ${\frak g}_h$ between themselves, then we obtain the boost ${\frak b}_{dSP}(x)={\frak g}_h(\omega t){\frak g}_p(\vec{x}) $ of the de Sitter-Painlev\'e chart $\{t,\vec{x}\}$ covering the same portion. Then  the canonical 1-forms, 
\begin{eqnarray}
\hat\omega^0&=&dt\,, \nonumber\\
\hat\omega^i&=&dx^i-\omega x^i dt \,,\label{1formdS}\\
\hat\omega^d&=&0\,,\nonumber
\end{eqnarray}
give the de Sitter-Painlev\'e metric   
\begin{equation}
ds^2=(1-\omega^2 \vec{x}^2)dt^2+2\omega \vec{x}\cdot d{\vec{x}}dt-d\vec{x}\cdot d\vec{x}\,,
\end{equation}
which shows that the event horizon of an observer staying at rest in $\vec{x}=0$ is the sphere $S^{d-2}$ of radius $|\vec{x}|=\omega^{-1}$ as in Fig. 1. 

{ \begin{figure}
    \centering
    \includegraphics[scale=0.45]{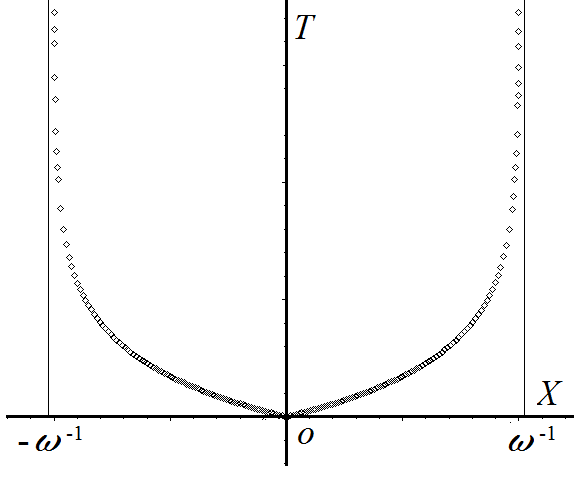}
    \caption{The null cone and the event horizons of an observer staying at $X=0$  in  the frame  with de Sitter-Painlev\'e coordinates $T, \,X$ }
  \end{figure}}

In this chart, the Killing vector fields ${\cal K}_{(AB)}=k_{(AB)}^{\mu}\partial_{\mu}$  that may have time-like domains \cite{Mau1,Mau2} are 
\begin{eqnarray}
&&{\cal K}_{(0d)}=-\frac{1}{\omega}\partial_t \,,\label{KH}\\
&&{\cal K}_{(0i)}=\nonumber\\
&&-e^{-\omega t}\left[x^i\partial_0+\left(\frac{\omega}{2}\,\vec{x}^2+\frac{1}{2\omega}(e^{2\omega t}-1)\right)\partial_i\right].\label{KK}
\end{eqnarray}
They are related to the generators of the natural representation carried by the space of scalar functions on $(M_{d},g)$. More specific, the energy operator, $H$, and  the generator of the Lorentz-like transformations, $K$, are defined as  \cite{C1}
\begin{equation}\label{HKKK}
H=-i\omega {\cal K}_{(0d)}\,, \quad K_i=-i{\cal K}_{(0i)}\,.
\end{equation}
The Lie algebra of this representation is completed by  the generators $J_{ij}=-i{\cal K}_{(ij)}$ of space rotations and  the Runge-Lenz type ones,   $R_{i}=-i{\cal K}_{(id)}$ \cite{C2,C1},  which will be not used here.

\section{Accelerating in two dimensions}

Now we have all the elements we need for presenting our proposal concerning the accelerated frames in de Sitter manifolds. We start with the simplest  two-dimensional case ($d=2$)  assuming that the coordinates $\{t,x\}$ of an accelerated frame in $(M_2,g)$ are introduced  by the specific boost
\begin{equation}\label{bacc}
{\frak b}_{acc}(t,x)= {\frak g}_r(\theta){\frak g}_h(\Omega t){\frak g}_r(-\theta){\frak g}_p(u)\,,
\end{equation}
where ${\frak g}_h$, ${\frak g}_p$ and ${\frak g}_r$ are the restrictions to $d=2$ of the $(d+1)$-dimensional matrices defined by Eqs. (\ref{gh}), (\ref{gp}) and respectively (\ref{gr}). These depend on the parameters 
\begin{eqnarray}
\Omega &=& \sqrt{\omega^2+a^2}\,,\label{Om}\\
\theta &=& \arctan \frac{a}{\omega}\,\label{theta},\\
u &=& \frac{1}{a}(e^{a x}-1)\,,\label{coordu}
\end{eqnarray}
that determine the geometry of the accelerated frame. This boost was defined in accordance with  the condition (I) obeying,
\begin{equation}\label{blim}
\lim_{a\to 0}{\frak b}_{acc}(t,x)={\frak b}_{dSP}(t,x)={\frak g}_h(\omega t){\frak g}_p(x)\,,
\end{equation}
since for $a\to 0$ we have $\Omega\to \omega$, $\theta\to 0$ and $u\to x$.

The boost (\ref{bacc}) gives rise to the following 1-forms
\begin{eqnarray}
\hat\omega^0&=&dt\, e^{ax} \,,\\
\hat\omega^1&=&dx\, e^{ax}-\frac{\omega}{2a}(e^{2ax}-1) dt\,, \\
\hat\omega^2&=&0\,.
\end{eqnarray}
determining the metric tensor of the accelerated chart $\{t,x\}$, 
\begin{eqnarray}
\!\!\!&&\!\!\!\!g(t,x)=\nonumber\\
\!\!\!&&\!\!\!\!\left(\begin{array}{cc}
e^{2ax}-\frac{\textstyle\omega^2}{\textstyle 4a^2}(e^{2ax}-1)^2& \frac{\textstyle\omega}{\textstyle 2a}e^{ax}(e^{2ax}-1)\\
 \frac{\textstyle\omega}{\textstyle 2a}e^{ax}(e^{2ax}-1)&-e^{2ax}
 \end{array}\right)\,,
\end{eqnarray} 
which is non-singular for any value of  $a\in {\Bbb R}$ since $\det(g)=-e^{4ax}\not= 0$. This  metric tensor fulfils automatically the condition (I) as long as the boost (\ref{bacc}) is set to satisfy Eq. (\ref{blim}). Then it is natural to obtain the de Sitter-Painlev\'e metric (for $d=2$) in the static limit,
\begin{equation}
\lim_{a\to 0}g(t,x)=\left(\begin{array}{cc}
1-\omega^2 x^2&\omega x\\
\omega x&-1
 \end{array}\right)\,.
\end{equation}
However, the condition (II) cannot be imposed {\em a priori} in a similar manner at the level of boost building  since there are no such  boosts in Minkowski's geometry. Therefore, we may feel  lucky recovering in the flat limit precisely the genuine Rindler metric,
\begin{equation}
\lim_{a\to 0}g(t,x)=\left(\begin{array}{cc}
e^{2ax}&0\\
0&-e^{2ax}
 \end{array}\right)\,.
\end{equation}
Thus we can say that the boost (\ref{bacc}) solves our problem correctly in accordance with the requirements (I) and (II), introducing the coordinates $\{t,x\}$ of an uniform accelerated frame. We say that  these are the {\em natural} coordinates of the accelerated frame  assuming that the accelerated observer stays at rest in $x=0$.

It remains to derive the concrete transformation between the accelerated frame defined above and a static frame supposed to have the de Sitter-Painlev\'e coordinates $\{T,X\}$. Therefore, we have to solve the equations resulted from the identity ${\frak b}_{dSP}(T,X)\,z_o={\frak b}_{acc}(t,x)\, z_o $. The solutions,
\begin{eqnarray}
T(t,x)\!\!\!&=&\!\!\!\frac{1}{\omega}\ln\left[\frac{1}{2}\frac{\omega^2}{\Omega^2}\left(\cosh \Omega t-1\right)\left(e^{2ax}+1\right)\right.\nonumber\\
&&+\left.\frac{\omega}{\Omega}\,e^{ax}\sinh\Omega t +1\right]\,,\\
X(t,x)\!\!\!&=&\!\!\!\frac{1}{2a}\left[-\frac{\omega^2}{\Omega^2}\left(\cosh\Omega t-1\right)\right.\nonumber\\
&&\left.+\frac{\omega}{\Omega}\sinh\Omega t\right]\left(e^{ax}-1\right)^2\nonumber \\
&& +\, \frac{e^{ax}}{a\Omega^2}\left(a^2 \cosh\Omega t + \omega^2\right)-\frac{1}{a}\,,\label{Xtx}
\end{eqnarray}
which represent the principal result of our proposal,  can be seen as the {\em Rindler-type} transformation in the case of the two-dimensional  de Sitter spacetimes.  This satisfies the requirements (I) and (II)   since $T(t,x)\to t$ and $X(t,x)\to x$ when $a\to 0$ while in the flat limit ($\omega \to 0$) we recover the genuine Rindler transformation as given by Eqs.  (\ref{T}) and (\ref{X}). Notice that the initial condition, $T(0,x)=0$ and $X(0,x)=u$, is the same as in the flat case. Moreover, the translation with $a^{-1}$ considered here can be removed in Eq.  (\ref{Xtx}) as in the genuine Rindler case of Eq. (\ref{X}) \cite{BD}.

On the other hand, these transformations help us to study how the conserved quantities of the static and accelerated frames are related among themselves. In terms of quantum observables we obtain the  identity
\begin{equation}\label{HHK}
H_{acc}=i\partial_t=H+a K\,,
\end{equation}
where $H$ and $K$ are the operators of the static frame as defined by Eqs. (\ref{HKKK})  for $d=2$.  Obviously,   Eq. (\ref{HHK}) is the same as in the flat case of the Rindler transformation continuous in $a=0$ we consider here. This fact is notable since  the  Minkowski  and  de Sitter observables denoted by $K$ have the same physical meaning,  but  very different structures.  

In general, other coordinates can be used for different purposes. For example, we can proceed as in the Rindler case, introducing another space coordinate,
\begin{equation}\label{phys}
x_{\parallel}=\frac{e^{ax}}{a}\in {\Bbb R}^+\,,
\end{equation}
which remains positive in the right-hand Rindler wedge (for $a>0$). This plays the role of a {\em  physical} space coordinate since the metric tensor of the chart $\{t,x_{\parallel}\}$ has the form
\begin{eqnarray}
\!\!\!&&\!\!\!\!g(t,x_{\parallel})=\nonumber\\
\!\!\!&&\!\!\!\!\left(\begin{array}{cc}
a^2 {x_{\parallel}}^2-\frac{\textstyle\omega^2}{\textstyle 4a^2}(a^2 {x_{\parallel}}^2-1)^2& \frac{\textstyle\omega}{\textstyle 2a}(a^2 x_{\parallel}^2-1)\\
 \frac{\textstyle\omega}{\textstyle 2a}(a^2 x_{\parallel}^2-1)&-1
 \end{array}\right)\,,
\end{eqnarray}  
(with $g_{11}=-1$). Notice that in this chart the metric tensor is singular in $a=0$ since $\det (g)=-a^2 x_{\parallel}^2$. 

When $a\not=0$ the chart $\{t,x_{\parallel}\}$  is useful for finding the equations of the 'light-cone' (or null cone) by integrating the equation $ds^2=0$. Of course, we have to look only for  solutions giving the worldlines passing through origin, $(0,a^{-1})$, since these are the borders of the future time-like domain of the observer at rest in $x_{\parallel \, 0} = a^{-1}$. After a few manipulations, we find the following equations
\begin{equation}
t=\frac{1}{\omega}\left(\ln \frac{\Omega-\omega\pm a}{\Omega+\omega\mp a}-\ln \frac{\Omega-a\omega x_{\parallel}\pm a}{\Omega+a\omega x_{\parallel}\mp a}\right)\,,
\end{equation} 
representing two worldlines that for $t\to\infty$ converge asymptotically to the event horizons whose positions are given by 
\begin{equation}\label{xppp}
x_{\parallel\, \pm}=\sqrt{\frac{1}{\omega^2}+\frac{1}{a^2}}\pm \frac{1}{\omega}\,.   
\end{equation}
More specific, the forward horizon (along the acceleration direction) is at $x_{\parallel\,+}$  and the back one at $x_{\parallel\,-}$. The interpretation is simple in terms of physical coordinates where we can measure the distances among the horizons and observer's position  $x_{\parallel\,0}$ which satisfies the natural condition $x_{\parallel \,-}<x_{\parallel\,0}<x_{\parallel\,+}$. We find that $x_{\parallel\,0}-x_{\parallel\,-}<x_{\parallel\,+} -x_{\parallel\,0}$ which shows how the acceleration pushes forward both these horizons without modifying their relative distance $x_{\parallel \,+}-x_{\parallel \,-}=2\omega^{-1}$ as in Fig. 2. A similar result can be obtained by using directly the coordinate (\ref{coordu}). Moreover, we observe that the natural coordinates of these horizons,
\begin{equation}
x_{\pm}=\frac{1}{a}\ln\left(\sqrt{1+\frac{a^2}{\omega^2}}\pm \frac{a}{\omega}\right)\,,
\end{equation}
have the expected limits $x_{\pm}\to \pm \omega^{-1}$ for $a\to 0$. 

{ \begin{figure}
    \centering
    \includegraphics[scale=0.45]{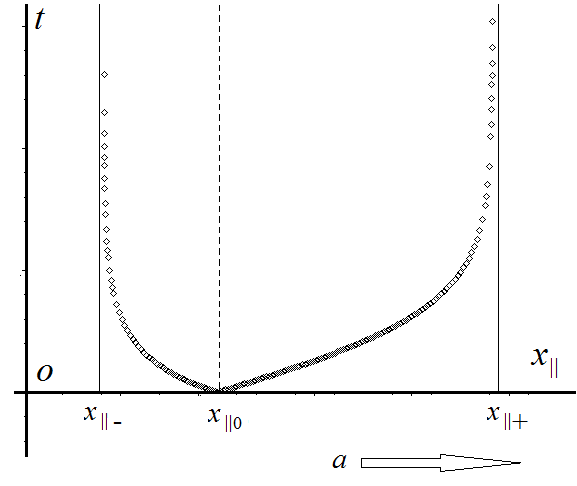}
    \caption{The null cone and the displaced horizons of an observer staying at rest in the accelerated frame at $x_{\parallel 0}=a^{-1}$. }
  \end{figure}}

\section{Accelerating in any dimensions}

The method of introducing coordinates using boosts allows us to generalize the definition of accelerated frames to any dimensions keeping the same structure of the boosts. Thus, supposing that the mobile frame of  coordinates $\{t,\vec{x}\}$ is accelerated in $(M_d,g)$ along to the axis $x^{d-1}$, we define the boosts  
\begin{equation}\label{baccd}
{\frak b}_{acc}(t,\vec{x}) = 
{\frak g}_r(\theta){\frak g}_h(\Omega t){\frak g}_r(-\theta)
{\frak g}_p(\vec{x})\,, 
\end{equation}
where $\vec{x}=(x^1,x^2,...x^{d-2},u)=(\vec{x}_{\perp},u)$ with $x^{d-1}=u$ as given by Eq. (\ref{coordu}) and the same parameters (\ref{Om}) and (\ref{theta}). We introduce thus the chart of natural coordinates $\{t,x^1,x^2,...x^{d-2},x\}$ whose canonical 1-forms
\begin{eqnarray}
&&\!\!\!\hat \omega^0 = e^{ax}dt\,,\\
&&\!\!\!\hat\omega^i = dx^i-\omega e^{ax} x^i dt\,, \quad i=1,2,..,d-2\,,\\
&&\!\!\!\hat\omega^{d-1}=e^{ax}dx+\left[\frac{a }{2}\,\vec{x}_{\perp}^2-\frac{1}{2a} \left(e^{2ax}-1\right)\right]\omega dt  \,, \\
&&\!\!\!\hat\omega^d = 0\,,
\end{eqnarray}
generate similar properties like in the two-dimensional  chart $\{t,x\}$, complying  with the requirements (I) and (II). Indeed, avoiding to write the complicated form of the metric tensor in this chart, we observe that for $a\to 0$ the above 1-forms become just the de Sitter-Painlev\'e ones (\ref{1formdS}), while in the flat limit ($\omega\to 0$) we obtain the familiar $d$-dimensional Rindler 1-forms
\begin{equation}
\lim_{\omega\to 0}~~\begin{array}{l}
\hat\omega^0\\
\hat\omega^i\\
\hat\omega^{d-1}\end{array}=\begin{array}{l}
e^{ax}dt\\
dx^i\\
e^{ax}dx
\end{array}\,.
\end{equation}

The transformations between the accelerated frame and the static one cannot be studied in the general case of any $d$, but analysing some concrete examples we can draw some general conclusions. First of all we can say that these have correct flat and static limits, satisfying thus (I) and (II).  Moreover,  it is worth pointing out that Eq. (\ref{HHK}) holds in any dimensions (but with $K_{d-1}$ instead of $K$).  This means that, in general,  for an accelerated frame in an arbitrary direction,  the energy operator of the accelerated frame can be expressed in terms of static observables as $H_{acc}=H+\vec{a}\cdot \vec{K}$.

In applications it is convenient to use the chart $\{t,x^1,x^2,...x^{d-2},x_{\parallel}\}$ depending on the physical coordinate $x_{\parallel}$ defined by Eq. (\ref{phys}). Then the canonical 1-forms become,
\begin{eqnarray}
&&\hat \omega^0 = a x_{\parallel} dt\,,\\
&&\hat\omega^i = dx^i-a\omega x_{\parallel} x^i dt\,, \quad i=1,2,..,d-2\,,\\
&&\hat\omega^{d-1}=dx_{\parallel}-h dt \,, \\
&&\hat\omega^d = 0\,,
\end{eqnarray}
where
\begin{equation}
h=\frac{\omega}{2a}[a^2(x_{\parallel}^2- \vec{x}_{\perp}^2)-1]\,.
\end{equation}
If we denote, in addition, $f=1-\omega^2 \vec{x}_{\perp}^2$,  we can write  the metric tensor in a simpler form,
\begin{eqnarray}
\!\!\!\!\!&&\!\!\!\!g(t,x^1,...,x_{\parallel})=\nonumber\\
\!\!\!\!\!&&\!\!\!\!\left( \begin{array}{ccccc}
a^2x_{\parallel}^2 f-h^2&\omega a x_{\parallel} x^1&\omega a x_{\parallel} x^2&\cdots&h\\
\omega a x_{\parallel} x^1&-1&0&\cdots&0\\
\omega a x_{\parallel} x^2&0&-1&\cdots&0\\
\vdots&\vdots&\vdots&&\vdots\\
h&0&0&\cdots&-1
\end{array}\right)\,,
\end{eqnarray}
but which is singular in $a=0$ having $\det (g)= -a^2 x_{\parallel}^2$ as in the two-dimensional case.

In this chart we can study the form of the event horizon of the accelerated observer (along to the axis $x^{d-1}$) located at $\vec{x}_0=(0,0,0,...a^{-1})$. Using the same method as in the two-dimensional case, we find that this remains a sphere $S^{d-2}$ of radius $\omega^{-1}$  having the center in $\vec{x}_c=(0,0,...,x_{\parallel\,c})$ with 
\begin{equation}
x_{\parallel\,c}=\frac{1}{2}(x_{\parallel\,+}+x_{\parallel\,-})= \frac{\Omega}{a\omega} >\frac{1}{a}\,,
\end{equation} 
as it results from Eq. (\ref{xppp}). In other words, the acceleration pushes forward the event horizon in an eccentric position  but without modifying its shape. 

\section{Concluding remarks}

We conclude that our definition of  accelerated frames on de Sitter spacetimes  seems to lead to a new geometric conjecture with a reasonable physical meaning. It would be interesting to compare our results with those reported so far  in the literature \cite{PG,BK1,BK1a,BK2,BTW}. 

The Nachtmann boosting method we applied here is very effective for introducing coordinates in de Sitter manifolds and pointing out geometrical properties. Therefore, we do not understand why this method is less used or quite ignored in investigating  physical systems in this geometry. We hope that the examples presented here will  turn people's attention  to the opportunities offered by this group theoretical approach. 

Thanks to this method,  we obtained accelerated charts  covering  the expanding portions of the de Sitter manifolds  that can be interpreted as expanding universes. Similar results may be obtained on the collapsing portions by applying the antipodal transformation. We hope that our proposal presented here could be the starting point for studying the motion of classical particles or even the quantum modes in uniform accelerated frames on the de Sitter expanding or collapsing universes. A crucial topic here could be the study of the Unruh effect in this geometry.


\subsection*{Acknowledgments}

This work was supported by a grant of the Romanian National Authority for Scientific Research. Program for research: Space Technology and Advanced Research  (STAR); project number: 72/29.11.2013 of the Romanian Space Agency.

\end{document}